\documentclass[12pt]{article}
\usepackage{amsfonts,epsfig,latexsym,amscd,amsmath,theorem,mathrsfs,color}

\graphicspath{{./images/}}

\newcommand{\R}{{\mathbb{R}}}

\newcommand{\beq}{\begin{equation}}
\newcommand{\eeq}{\end{equation}}
\newcommand{\bea}{\begin{eqnarray}}
\newcommand{\eea}{\end{eqnarray}}
\newcommand{\ben}{\begin{eqnarray*}}
\newcommand{\een}{\end{eqnarray*}}
\newcommand{\bem}{\begin{enumerate}}
\newcommand{\eem}{\end{enumerate}}

\newcommand{\ra}{\rightarrow}

\newcommand{\cd}{\partial}

\newcommand{\wh}{\widehat}
\newcommand{\less}{\backslash}

\newcommand{\ignore}[1]{}

\newcommand{\mvec}{\mbox{\boldmath{$m$}}}
\newcommand{\xvec}{\mbox{\boldmath{$x$}}}

\newcommand{\uvec}{{\mbox{\boldmath{$u$}}}}
\newcommand{\vvec}{\mbox{\boldmath{$v$}}}

\newcommand{\evec}{\mbox{\boldmath{$e$}}}

\newcommand{\Avec}{\mbox{\boldmath{$A$}}}
\newcommand{\yvec}{\mbox{\boldmath{$y$}}}
\newcommand{\zvec}{\mbox{\boldmath{$z$}}}

\renewcommand{\phi}{\varphi}

\newcommand{\attract}{${\scriptscriptstyle \rm attract}$}
\newcommand{\repel}{${\scriptscriptstyle \rm repel}$}

\newcommand{\eqnew}{{\scriptscriptstyle \rm new}}
\newcommand{\eq}[1]{{\mbox{\boldmath{$\scriptstyle #1$}}}}

\theoremstyle{plain}

{\theorembodyfont{\rmfamily}

}

\newcommand{\news}{\setcounter{equation}{0}}

\definecolor{TW-color}{RGB}{100,0,100}
\definecolor{Error-color}{RGB}{250,0,0}

\begin{document}

\title{Intervortex forces in competing-order superconductors}
\author{Martin Speight\thanks{E-mail: {\tt j.m.speight@leeds.ac.uk}} and Thomas Winyard\thanks{E-mail: {\tt t.winyard@leeds.ac.uk}}\\
School of Mathematics, University of Leeds\\
Leeds LS2 9JT, England}

\maketitle


\begin{abstract}
The standard Ginzburg-Landau model of competing-order superconductors, applicable to various high $T_c$ cuprates, is studied. It is observed that this model possesses two distinct species of vortex, and consequently has two distinct integer valued topological charges. A simple point particle model of long range forces between (anti)vortices of any species is developed and compared with numerical simulations of the full field theory, excellent agreement being found. Some of the results are quite counterintuitive. For example, a parameter regime exists where vortices of one species repel both vortices and antivortices of the other.
\end{abstract}


\news

\section{Introduction}\news\label{sec:intro}

High $T_c$ cuprate superconductors often exhibit a superconducting ground state that is in close proximity to other ordered ground states. The standard approach models these two phases separately with separate order parameters. However, it has been shown that when in close proximity the superconducting state competes with these other orders, for example anti-ferromagnetic order \cite{aroberkalzha,lake2001spins} or charge order \cite{chang2012direct,wu}. In particular there has been considerable recent interest in such models, driven by experimental results, showing the importance of charge order in underdoped cuprates \cite{gabovich2010competition,chang2012direct,grissonnanche2014direct,fradkin2015colloquium}. 

If a magnetic field is applied to such a system, vortices form, locally suppressing the superconducting state in the core. This leads to competing correlations in the core, studied both theoretically \cite{aroberkalzha,hu2002theory,gao2011model} and experimentally \cite{hoffman2002four,lake2001spins,curran2011vortex,machida2016bipartite} in cuprates. In addition it has also been shown that in $YBa_2Cu_3O_y$, vortex cores overlap before $H_{c_2}$ is reached, allowing charge order across the system \cite{wu}.

A common tool used to study competing phases, is extending the target space to include the competing order parameters. The extended target space comes with additional constraints, such that suppression of the dominant phase is matched by excitation of the competing phase. Historically this was introduced in cuprates to model the competition between the superconducting phase and anti-ferromagnetic phase as an $SO(5)$ model. The approach considered a coupled complex valued order parameter $\Delta$ for the superconducting phase, and a vector valued order parameter $\mvec=(m_1,m_2,m_3)$ for the antiferromagnetic phase, and phase competition introduced through the constraint $|\Delta|^2+|\mvec|^2=\mbox{const}$ \cite{aroberkalzha}. Hence the composite order parameter $(\Delta,\mvec)$ takes values in a 4-dimensional sphere inside $\R^5$.

Recently it has been proposed that a similar approach using an $SO(3)$ model, where the target space is expanded to a two-sphere $S^2 \subset \mathbb{R}^3$, can be used to model the competition between superconductivity and charge order \cite{karmengan}. Restricting a half filled attractive Hubbard model to nearest neighbour hopping leads to the superconducting and charge density wave orders becoming degenerate in energy. This suggests an $S^2$ order parameter \cite{yang1989eta,zhang1990pseudospin,yang1990so,burkov2008stability,ganesh2009collective}, formed of a superconducting component, written as the complex field $\Delta$, and charge density wave component, written as the real field $\rho$. These fields are subject to the constraint $|\Delta|^2 + \rho^2 = c^2$, such that $|\rho|$ is maximal where $|\Delta|$ vanishes, and vice versa. Hence, we will assume that the north pole ($\rho = c$) and south pole ($\rho = -c$) of the $S^2$ target space correspond to two different charge density wave orders (with different dominating sub-lattices), while the equator ($\rho = 0$) exhibits the $U(1)$ superconducting phase.

Note that there have been studies of such competing phase models in uncharged systems \cite{efetov2013pseudogap,hayward2014angular,wachtel2014transverse,moor2014topological}, but such systems do not admit finite energy vortex solutions. As we are interested in vortices in this paper we will deal entirely with the charged model.

We also note that the effect of competing order is of general interest in superconductivity. It is important to understand such systems and their generalizations, with a focus on their solitonic excitations, from multi-component systems with density-density couplings \cite{garaud2015vortex} (which also exhibits non-trivial vortex interactions) to competition with spin density waves \cite{vavilov2010coexistence,vorontsov2010superconductivity}.

This paper will focus on the continuous effective Ginzburg-Landau (GL) formalism proposed in \cite{karmengan,sarkargan}, which is derived from the attractive $SO(3)$ Hubbard model mentioned above. It is similar to other phenomenologically proposed models \cite{meier2013effect,caplan2015long,caplan2017dimensional}, introduced in an attempt to model the experimental observation of competing phases in charged systems. To derive a GL model one must take the Hubbard model and assume that anisotropy near the charge ordered, superconducting transition is negligible. Taking this isotropic limit and assuming a quadratic symmetry breaking term gives fields subject to the free energy density,
\beq\label{Edef}
{\cal E}=\frac\chi2\left|(\nabla-\frac{2ie}{\hbar}\Avec)\Delta\right|^2+\frac{1}{8\pi}|\nabla\times\Avec|^2+\frac\chi2|\nabla\rho|^2-|\Delta|^2-(1-\delta)\rho^2,
\eeq
where $\Avec$ is the electromagnetic gauge potential coupled to the charged field $\Delta$, and $\chi$ and $\delta$ are positive constants, $\delta$ representing the strength of next-nearest neighbour hopping. As with the Hubbard model, order competition is imposed via the constraint $|\Delta|^2+\rho^2=c^2$. In mathematical terms, this is an example of a gauged sigma model, objects of strong intrinsic interest.

The purpose of this paper is to develop a theory of the long range interactions between vortices in this model within the point vortex formalism. A key observation is that the model supports two distinct species of vortex which we call North vortices, with $\rho=c$ in the vortex core, and South vortices, with $\rho=-c$ in the vortex core, and that each of these has an antivortex counterpart (possessing a quantum of {\em negative} magnetic flux).

While vortices have been previously studied \cite{karmengan,sarkargan,wachtel2015signatures}, there has been no detailed study of the different (anti)vortex interactions in the model. One paper briefly considered the effect that introducing charge order has on the strength of purely superconducting interactions in the Hubbard model\cite{karmakar2018vortex}. In that paper, vortex interactions were approximated to be that of a strongly type II single component model, with a numerically motivated exponential correction term, dependent on the value of $\delta$. However, in this paper we demonstrate that to understand the interactions, one cannot separate the superconducting and charge order components and treat them separately. We will also show that the interactions act as Bessel functions. Our detailed study of the interactions in the model will lead to a typology argument, similar to the standard single component Ginzburg-Landau model, but with additional complications due to the multiple species of vortex.

Since the model supports two different species of vortex, it possesses {\em two} integer-valued topological charges: the total number $n$ of magnetic flux quanta, and the half-degree $d$ of the map $\R^2\ra S^2$ defined by $(x_1,x_2)\mapsto({\rm Re} \Delta(x_1,x_2), {\rm Im} \Delta(x_1,x_2), \rho(x_1,x_2))/c$, or, equivalently the net numbers of North vortices $k_+$ and South antivortices $k_-$. This pair of integers cannot change under any smooth deformation of the fields $\Delta,\rho,A_i$ preserving finite total energy.

 We will see that the interaction between (anti)vortices of all types depends crucially on the 
coupling parameter
\beq\label{mudef}
\mu=\frac{\hbar\delta}{2\sqrt{2\pi}e\chi c}
\eeq
which plays a role analogous to the Ginzburg-Landau parameter in conventional (single component) GL theory. If $\mu>1$, vortices of any species repel one another, as do antivortices of any species, while vortices always attract antivortices. If $\mu<1$, the behaviour is more surprising: like vortices attract, as do like antivortices, but {\em unlike} vortices repel, as do unlike antivortices, {\em and} unlike vortex-antivortex pairs. The regime of critical coupling $\mu=1$ is particularly subtle with various combinations of vortices and antivortices experiencing no static interactions at all. The situation is summarized in table \ref{table1}. This constitutes the equivalent of the familiar typology argument for the standard single component Ginzburg-Landau model, where the parameter $\mu$ is now the Ginzburg-Landau parameter, determining the interaction type. Hence for $\mu > 1$ we call this a type II superconductor, for $\mu < 1$ a type I superconductor and $\mu = 1$ a critically coupled superconductor.

\begin{table}[htb]
\begin{center}
\begin{tabular}{ccc}
$\mu<1$ & $\mu=1$ & $\mu>1$ \\
\begin{tabular}{c|cccc}
		& $N$ 		& $\bar{N}$ 	& $S$ 		& $\bar{S}$  \\ \hline
$N$ 		& \attract	& \attract	& \repel	& \repel \\
$\bar{N}$	&		& \attract	& \repel 	& \repel \\
$S$		&		& 		& \attract	& \attract \\
$\bar{S}$	&		&		&		& \attract \\ 
\end{tabular}&

\begin{tabular}{c|cccc}
		& $N$ 		& $\bar{N}$ 	& $S$ 		& $\bar{S}$ \\ \hline 
$N$ 		& 0		& \attract	& \repel		& 0 \\
$\bar{N}$	&		& 0		& 0	 	& \repel \\
$S$		&		& 		& 0		& \attract \\
$\bar{S}$	&		&		&		& 0 \\ 
\end{tabular}&
\begin{tabular}{c|cccc}
		& $N$ 		& $\bar{N}$ 	& $S$ 		& $\bar{S}$ \\ \hline
$N$ 		& \repel	& \attract	& \repel	& \attract \\
$\bar{N}$	&		& \repel	& \attract 	& \repel \\
$S$		&		& 		& \repel	& \attract \\
$\bar{S}$	&		&		&		& \repel \\ 
\end{tabular}

\end{tabular}
\caption{Summary of interactions between (anti)vortex pairs. $N$ denotes North vortex, $S$ denotes South vortex and an overbar denotes the corresponding antivortex. The $0$ entries in the $\mu=1$ table indicate (anti)vortex pairs which experience no interaction: their total energy is independent of their separation.}
\label{table1}
\end{center}
\end{table}

The rest of this paper is structured as follows. In section \ref{sec:vortices} we choose length, energy and charge units to reduce the GL model to a standard gauged sigma model, review its topological properties, and construct its (anti)vortices, paying particular attention to their asymptotics at spatial infinity. In section \ref{sec:pointvortex} we develop a theory of long range intervortex interactions by modelling vortices as solutions of the linearization of the sigma model about its vacuum, in the presence of appropriate point sources at the vortex centre, chosen to replicate the vortex's large $r$ behaviour. This models vortices as composite point particles carrying a scalar monopole charge, inducing a real scalar field of mass $\mu$ (roughly, the field $\rho$) and a magnetic dipole moment inducing a vector field
of mass $1$ (roughly, $A_i$). The interaction energy between pairs of such point particles is easily computed, producing the predictions of table \ref{table1}, as well as precise asymptotic formulae for the interaction energies valid at large separation. In section \ref{sec:numerics} we verify these predictions by numerically computing the interaction energy of (anti)vortex pairs via a gradient descent energy minimization method. Finally, section \ref{sec:conc} presents some concluding remarks.

\section{Competing-order vortices}\news\label{sec:vortices}

We first choose scales to minimize the number of parameters in the free energy \eqref{Edef}. Let 
\bea
{\cal E}&=&\lambda_{\cal E}{\cal E}^\eqnew-c^2,\quad
x_i=\lambda_xx^\eqnew_i,\quad 
A_i=\lambda_AA_i^\eqnew,\nonumber \\
(u_1+iu_2,u_3)&=&(\Delta/c,\rho/c),\nonumber \\
D_i\uvec&=&\frac{\cd\uvec}{\cd x_i^\eqnew} - A_i^\eqnew\evec\times\uvec,
\eea
 where $\evec=(0,0,1)$.
Then, with the choices
\beq
\lambda_{\cal E}=4\pi\chi^2 c^4\left(\frac{2e}{\hbar}\right)^2,\quad
\lambda_x^2=\frac{\chi c^2}{\lambda_{\cal E}},\quad
\lambda_A=\frac{\hbar}{2e\lambda_x},
\eeq
we find that
\beq
{\cal E}^\eqnew=\frac12 D_i\uvec\cdot D_i\uvec+\frac12 (B^\eqnew)^2+\frac{\mu^2}2(\evec\cdot\uvec)^2
\eeq
where $B^\eqnew=\cd_1^\eqnew A_2^\eqnew-\cd_2^\eqnew A_1^\eqnew$ and $\mu$ is defined in equation \eqref{mudef}. We henceforth discard the superscript ``${\rm new}$.''

 The total energy of
a pair of fields $(\uvec, A)$ is the integral
\beq
E=\int_{\R^2}{\cal E} dx_1 dx_2.
\eeq
In order for this to be finite, $\uvec$, at large $r$ (where $(x_1,x_2)=:r(\cos\theta,\sin\theta)$), must approach the equator $u_3=0$ on $S^2$. It need not, however, be constant: it may wind around the equator
\beq
\uvec\sim (\cos n\theta,\sin n\theta,0)
\eeq
some integer $n$ times. Then finite energy also implies $|D\uvec|\sim 0$ as $r\ra\infty$, so $\Avec\sim \frac{n}{r}(-\sin\theta,\cos\theta)$, whence, by a standard application of Stokes's Theorem one finds that the total magnetic flux of any finite energy configuration is quantized,
\beq
\int_{\R^2}Bdx_1dx_2=2\pi n.
\eeq

\begin{figure}
\center
\begin{tabular}{cc}
	\includegraphics[width=0.9\linewidth]{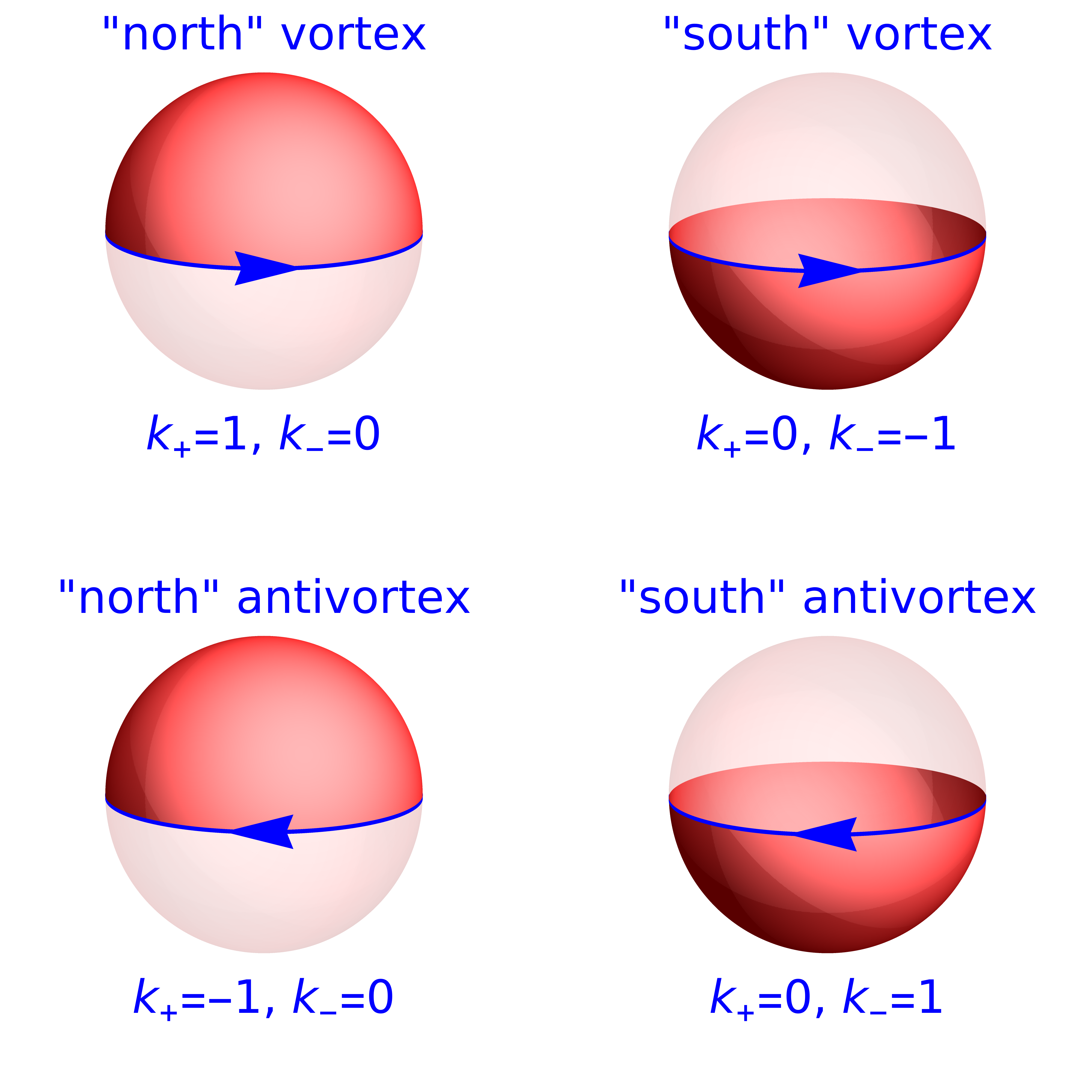} 
\end{tabular}
	\caption{The field values attained by the four species of (anti)vortex. The field $\uvec(\xvec)$ wraps the circle at spatial infinity once around the equator in the direction indicated, anticlockwise for vortices, clockwise for antivortices (viewed from above the North pole). The (anti)vortex interior then covers either the Northern or the Southern hemisphere once. The topological charges $k_+,k_-$ measure the number of times the field assumes the pole values $(0,0,1)$ and $(0,0,-1)$ respectively, counted with orientation and multiplicity. These poles indicate two different sub-lattices for the charge density wave order in the core of the (anti)vortex.}
	\label{Fig:vortex_picture}
\end{figure}

If $n\neq 0$, there must be points in the plane where $u_1+iu_2=0$. Note, however, that these come in two distinct species since $u_3$ may take the value $+1$ or $-1$ at each such point. Consider a point $\xvec^+$ where $\uvec(\xvec^+)=(0,0,1)$. This point itself may be assigned a sign $\sigma(\xvec^+)$ according to whether the field $\uvec(\xvec)$ is locally an orientation preserving ($\sigma=+1$) or orientation reversing ($\sigma=-1$) map close to
$\xvec^+$. The sum of these signs over all points where $\uvec=(0,0,1)$ is an integer-valued topological invariant of the field $\uvec$,
\beq
k_+=\sum_{\eq{x}\in \eq{u}^{-1}(\eq{e})}\sigma(\xvec)
\eeq
which we may interpret as the net excess of North vortices over North antivortices in the field configuration. We may similarly assign a sign $\sigma(\xvec^-)$ to each point $\xvec^-$ in the plane at which $\uvec(\xvec^-)=(0,0,-1)$. Again, $\sigma(\xvec^-)=+1$ if $\uvec(\xvec)$ is locally orientation preserving and $\uvec(\xvec^-)=-1$ if it is locally orientation reversing. One should note, however, that, while $(u_1,u_2)$ is a good oriented local coordinate system for $S^2$ in a neighbourhood of $(0,0,1)$, it is {\em anti-oriented} in a neighbourhood of $(0,0,-1)$, so each point with $\sigma(\xvec^-)=+1$ contributes {\em negatively} to the winding of the field $\uvec$ about the equator in $S^2$. Hence, the integer-valued topological invariant associated with the South (anti)vortex positions 
\beq
k_-=\sum_{\eq{x}\in \eq{u}^{-1}(-\eq{e})}\sigma(\xvec)
\eeq
represents the net excess of South {\em antivortices} over South {\em vortices} in the field configuration. One sees that the winding number at spatial infinity, which determines the total magnetic flux, is determined by $k_+,k_-$ as
\beq
n=k_+-k_-.
\eeq
Furthermore, the total signed area in $S^2$ covered by the mapping $\uvec(\xvec)$ is $2\pi(k_++k_-)$, so we may identify $k_++k_-$ has the {\em half-degree} of the map $\uvec:\R^2\ra S^2$.
The four types of (anti)vortex supported by this model are summarized pictorially in Figure \ref{Fig:vortex_picture}. We reiterate that the difference between North and South vortices is the dominant sub-lattice for the charge density wave order in the core of the vortex.

To understand the (anti)vortices in more detail, we must numerically solve the Euler-Lagrange equations for the functional $E$,
\bea
P_\eq{u}(-D_iD_i\uvec+\mu^2(\evec\cdot\uvec)\evec)&=&0,\\
-\cd_i\cd_iA_j+\cd_j\cd_iA_i-\evec\cdot(\uvec\times D_i\uvec)&=&0,
\eea
where $P_\eq{u}$ denotes projection orthogonal to $\uvec$, that is, $P_\eq{u}(\vvec):=\vvec-(\uvec\cdot\vvec)\uvec$. These are consistent with the ansatz
\bea
\uvec^N&=&(\sin f(r)\cos\theta, \sin f(r)\sin\theta,\cos f(r))\\
\Avec^N&=&\frac{a(r)}{r}(-\sin\theta,\cos\theta)
\eea
where the profile functions $f$, $a$, satisfy the coupled ODE system
\bea
f''+\frac{1}{r}f'-\frac{(1-a)^2}{r^2}\sin f\cos f+\mu^2\sin f\cos f&=&0\label{ode1}\\
a''-\frac{1}{r}a'+\sin^2f(1-a)&=&0\label{ode2}
\eea
subject to the boundary conditions $f(0)=a(0)=0$, $f(\infty)=\pi/2$, $a(\infty)=1$. Having found $f$ and $a$, we may easily construct the other three
species of (anti)vortex,
\bea
\uvec^{S}&=&(\sin f(r)\cos\theta, \sin f(r)\sin\theta,-\cos f(r)),\qquad \Avec^{S}=\frac{a(r)}{r}(-\sin\theta,\cos\theta),\nonumber\\
\uvec^{\bar{N}}&=&(\sin f(r)\cos\theta, -\sin f(r)\sin\theta,\cos f(r)),\qquad \Avec^{\bar{N}}=\frac{a(r)}{r}(\sin\theta,-\cos\theta),\nonumber\\
\uvec^{\bar{S}}&=&(\sin f(r)\cos\theta, -\sin f(r)\sin\theta,-\cos f(r)),\qquad \Avec^{\bar{S}}=\frac{a(r)}{r}(\sin\theta,-\cos\theta).
\label{menagerie}
\eea

\begin{figure}
\center
	\includegraphics[width=1\linewidth]{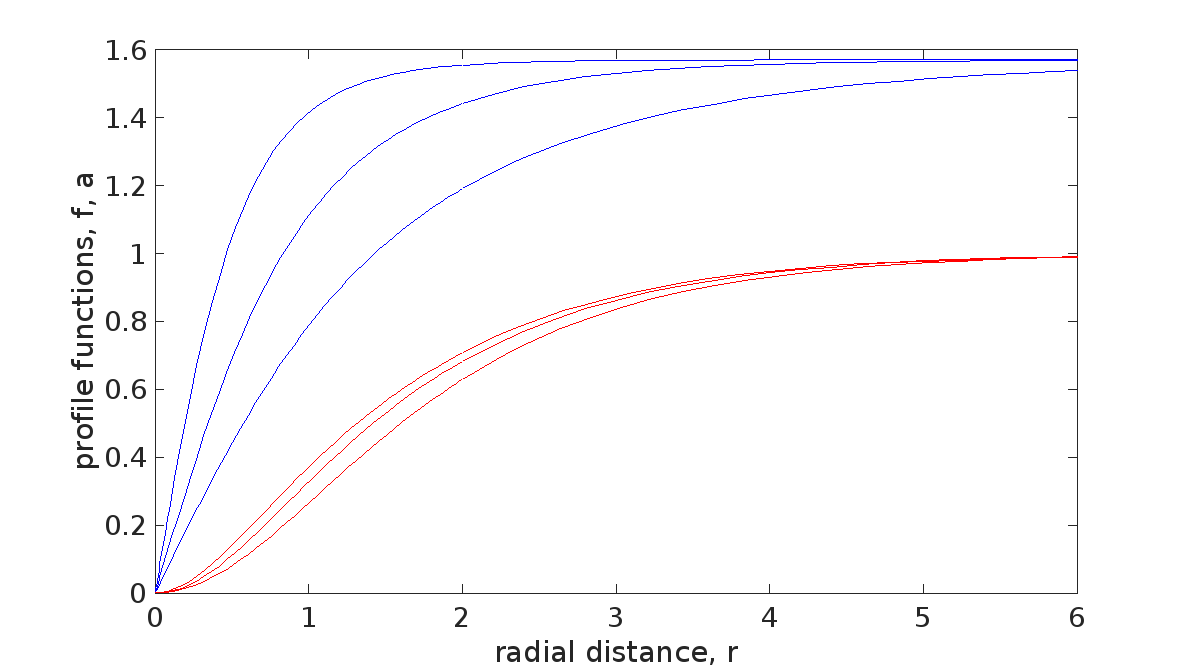} 
	\caption{The profiles functions $f(r)$ (blue curves) and $a(r)$ (red curves) of a North vortex at couplings $\mu=2$ (top), $\mu=1$ (middle) and
$\mu=0.5$ (bottom).}
	\label{Fig:profiles}
\end{figure}

\begin{figure}
\center
	\includegraphics[width=1\linewidth]{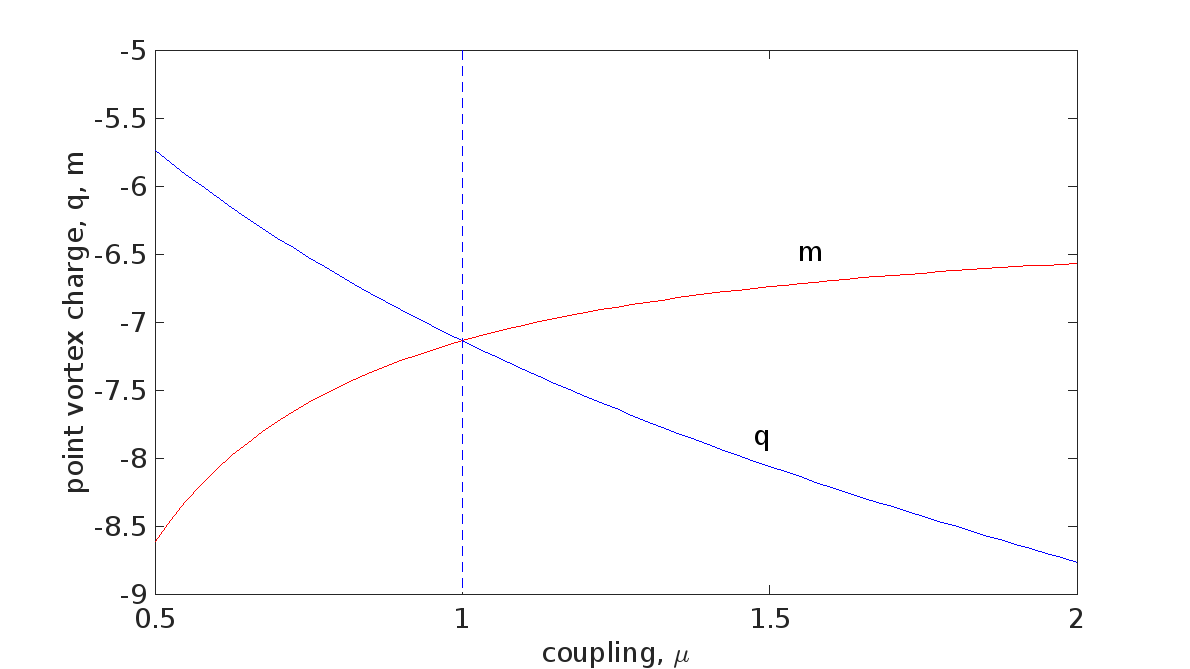} 
	\caption{The large $r$ shooting parameters $q,m$ of the North vortex solution as functions of the coupling $\mu$. These may be interpreted as the scalar monopole charge ($q$) and magnetic dipole moment ($m$) of the corresponding point vortex.}
	\label{Fig:charges}
\end{figure}

The system \eqref{ode1}, \eqref{ode2} does not appear to be integrable, so we resort to numerical integration to find $f,a$. Regularity at the origin requires $f(r)\sim\alpha_1 r$ and $a(r)\sim \alpha_2 r^2$ for some constants $\alpha_1,\alpha_2$. For large $r$, $\wh{f}(r):=f(r)-\pi/2$ and
$\wh{a}(r):=a(r)-1$, being small, should be asymptotic to decaying solutions of the {\em linearization} of the system about $(f,a)=(\pi/2,1)$,
\bea
\wh{f}''+\frac{1}{r}\wh{f}'-\mu^2\wh{f}&=&0, \label{lode1}\\
\wh{a}''-\frac1r\wh{a}'-\wh{a}&=&0. \label{lode2}
\eea
Hence, at large $r$,
\beq
f(r)\sim \frac{\pi}{2}+\frac{q}{2\pi}K_0(\mu r),\qquad a(r)\sim 1+\frac{m}{2\pi}rK_1(r),
\eeq
where $K_0,K_1$ are modified Bessel's functions of the second kind, and $q,m$ are unknown constants. The factors of $2\pi$ are included for later convenience. Our numerical strategy is to solve \eqref{ode1}, \eqref{ode2} on $[r_0,R]$, with $r_0>0$ small and $R$ large by shooting rightwards from $r_0$, using  $(\alpha_1,\alpha_2)$ as shooting parameters, leftwards from $R$ using $(q,m)$ as shooting parameters, and imposing that $f,a$ and their derivatives match at some interior point $r_1$ of order $1$. The results of this scheme for various values of the coupling $\mu$ are depicted in  
Figure \ref{Fig:profiles}. Of particular interest are the values of the constants $(q,m)$ as functions of $\mu$, depicted in Figure \ref{Fig:charges}.
Note that $q\equiv m$ when $\mu=1$. This is not a coincidence: the system \eqref{ode1}, \eqref{ode2} reduces to a {\em first} order system at this critical value of the coupling,
\beq
f'=\frac{1-a}{r}\sin f, \qquad a'=r\cos f,
\eeq
from which it follows immediately that $q\equiv m$. This is a symptom of the {\em self duality} (or BPS property) enjoyed by the model at $\mu=1$, whose full consequences are both deep and far ranging \cite{yan,romspe}. In this paper we will concentrate on the case $\mu\neq 1$, however.

\section{The point vortex model}\news\label{sec:pointvortex}

It is convenient to think of (anti)vortices as static solutions of the Lorentz invariant model on $(2+1)$-dimensional Minkowski space whose static energy is $E$, that is, the model with Lagrangian density
\beq\label{relmod}
{\cal L}=\frac12D_\mu\uvec\cdot D^\mu\uvec-\frac14 F_{\mu\nu}F^{\mu\nu}-\frac{\mu^2}{2}(\evec\cdot\uvec)^2,
\eeq
where $F_{\mu\nu}=\cd_\mu A_\nu-\cd_\nu A_\mu$, spacetime indices $\mu,\nu$ run over $0,1,2$, and the Minkowski metric has signature $+--$. We have merely extended the indices to include time components for all derivatives and the gauge field.
We emphasize that this is a mathematical device, introducing 2nd order Lorentzian dynamics. This allows us to access some techniques and results familiar in the study of topological solitons in high energy physics. We certainly do {\em not} assert that the time dynamics defined by this relativistic extension is relevant to competing order superconductors.

The key observation is that static vortices, far from their core, are indistinguishable from solutions of the linearization of the model \eqref{relmod} about the vacuum (meaning $A_\mu=0$, $\uvec=(1,0,0)$) in the presence of appropriate point sources placed at the vortex centre. Since physics is model independent, the forces between well-separated vortices should coincide with those between the corresponding point sources interacting via the fields they induce in the linear theory. These are easily computed, yielding an asymptotic formula for the interaction energy between well-separated vortices. This underlying idea was introduced by Manton to study long-range forces between magnetic monopoles \cite{man}, and subsequently applied to nuclear Skyrmions by Schroers \cite{sch}. It was adapted to vortices in the conventional Ginzburg-Landau model in \cite{spe-long_range}, then multicomponent vortices in \cite{babcarspe,carbabspe,chakommilandfarpee}. 

Our first task is to identify the point sources that replicate the vortex asymptotics, and to do this we must first re-write it in the gauge in which,
as $r\ra\infty$, $\uvec\ra (1,0,0)$ in every direction, that is, the gauge where $u_2=0$ and $u_1\geq 0$. This is accomplished by applying the singular (at $r=0$) gauge transformation $(u_1+iu_2,u_3)\mapsto (e^{-i\theta}(u_1+iu_2),u_3)$. The order parameter takes the form $\uvec=(\cos\Theta,0,\sin\Theta)$ in this gauge, the vacuum is $\Theta=0$ and the North vortex has
\bea
\Theta(r)&=&f(r)-\frac\pi2\sim \frac{q}{2\pi}K_0(\mu r),\nonumber \\
(A_0,A_1,A_2)&=&\frac{a(r)-1}{r}(0,-\sin\theta,\cos\theta)\sim \frac{m}{2\pi}(0,\cd_2,-\cd_1)K_0(r).
\eea
These are precisely \cite{spe-long_range} the fields induced in the linearized model
\beq
{\cal L}_{lin}=\frac{1}{2}\cd_\mu\Theta\cd^\mu\Theta-\frac{\mu^2}{2}\Theta^2+\rho\Theta-\frac14 F_{\mu\nu}F^{\mu\nu}+\frac12 A_\mu A^\mu +j_\mu A^\mu
\eeq
by the static sources
\beq
\rho=q\delta(\xvec), \qquad
(j_0,j_1,j_2)=m(0,\cd_2,-\cd_1)\delta(\xvec),
\eeq
so our linearized model of a North vortex is a composite point source consisting of a scalar monopole of charge $q^N=q$, inducing a real scalar field 
$\Theta$ of mass $\mu$, and a magnetic dipole of moment $m^N=m$ inducing a Proca field $A_\mu$ of mass $1$. The corresponding sources for the other species of (anti)vortex follow immediately by unwinding \eqref{menagerie}. All are scalar monopole/magnetic dipole composites, with charges
\beq
(q^N,m^N)=(q,m),\quad
(q^S,m^S)=(-q,m),\quad
(q^{\bar{N}},m^{\bar{N}})=(q,-m),\quad
(q^{\bar{S}},m^{\bar{S}})=(-q,-m).
\label{charges}
\eeq

The interaction Lagrangian for a pair of sources $(\rho^{(1)},j_\mu^{(1)})$, $(\rho^{(1)},j_\mu^{(1)})$ is
\beq
L_{int}=\int_{\R^2}(\rho^{(1)}\Theta^{(2)}+j_\mu^{(1)}A^\mu_{(2)})dx_1dx_2
\eeq
where $(\Theta^{(2)},A_\mu^{(2)})$ are the fields induced by the second source. We apply this in the case where the sources are static scalar monopole/magnetic dipole composites of charges $(q_{1},m_1)$, $(q_2,m_2)$ located at $\yvec$ and $\zvec$ respectively. The result is a function of
$s:=|\yvec-\zvec|$, the vortex separation. It may be interpreted as {\em minus} the interaction energy of the source pair, so
\beq\label{Eint}
E_{int}(s)=-L_{int}=\frac{1}{2\pi}\left[m_1m_2K_0(s)-q_1q_2K_0(\mu s)\right].
\eeq
If $\mu>1$, the first term, representing magnetic interactions, dominates at large $s$, whereas if $\mu<1$, the second term, representing scalar
interactions dominates. By choosing $(q_1,m_1)$, $(q_2,m_2)$ from the list \eqref{charges}, we obtain long range interaction energies between (anti)vortices of any species. The nature of these interactions is summarized in Table \ref{table1}. The zero entries for critical coupling, $\mu=1$,
follow from the observation that $q=m$ here. Our calculation establishes that the leading order interactions for $NN$, $SS$, $N\bar{S}$ and $S\bar{N}$
pairs vanish in this case. In fact, the self-duality structure can be used to prove that the interaction vanishes exactly for these pairs \cite{yan}: static solutions exist with the individual vortices placed at any points in the plane when $\mu=1$.

Of course, these predicted interaction potentials are based on a leap of faith -- that physics is model independent. This particular faith allows, indeed encourages, scepticism in its acolytes. Luckily it also admits a definitive test: we can compute the energy between vortices held at a fixed separation by numerical simulation of the original nonlinear model. This is the subject of the next section.

\section{Numerical results}\news\label{sec:numerics}

How can we compute the interaction energy $E^{NN}_{int}(s)$ between two North vortices held distance $s$ apart? Note that no such {\em static} solution
exists (unless $s=0$, or $\mu=1$), precisely because vortices exert forces on one another. The answer is that we solve a {\em constrained} minimization problem for the energy functional $E$: we minimize among all fields having $k_+=2$ and $k_-=0$ subject to the constraint that 
$\uvec(s/2,0)=\uvec(-s/2,0)=\evec$. In practice, we discretize space, replacing spatial derivatives by difference operators on a regular $n_1\times n_2$ lattice of spacing $h$ (we used $n_1=n_2=251$ and $h=0.1$). This replaces the
continuum energy functional $E(\uvec,\Avec)$ by a discrete approximant $E_{dis}:{\cal C}_{dis}\ra\R$ where ${\cal C}_{dis}=(S^2)^{n_1n_2}\times (\R^2)^{n_1n_2}$ is the discretized configuration space. We then construct an appropriate initial guess $\uvec_{i,j}$, $\Avec_{i,j}$ with, around the boundary of the lattice, $\uvec_{i,j}\cdot\evec=0$ and winding $2$, and 
\beq
\uvec_{\pm i_0,0}=\evec,
\label{constraint} 
\eeq
where
$s=2i_0h$. We then minimize $E_{dis}$ among all points in ${\cal C}_{dis}$ satisfying the constraint \eqref{constraint} using arrested Newton flow 
\cite{spewin} for the
function $E_{dis}$, but {\em never updating} $\uvec_{\pm i_0,0}$ (or $\uvec$, $\Avec$ on the boundary of the lattice). This automatically produces a solution of the Euler-Lagrange equations for our energy functional on $\R^2\less\{(\pm s/2,0)\}$ satisfying the constraint \eqref{constraint} at the missing points. An alternative to this procedure is to solve the Euler-Lagrange equations on $\R^2\less\{(\pm s/2,0)\}$ directly, an approach exploited for the standard GL model in \cite{chapeefarmil}. Having computed the lowest energy among all $(k_+,k_-)=(2,0)$ field configurations with $\uvec(\pm s/2,0)=\evec$, we then subtract twice the energy of a single North vortex to obtain
$E_{int}^{NN}(s)$. 

\begin{figure}
\center
	\includegraphics[width=1.0\linewidth]{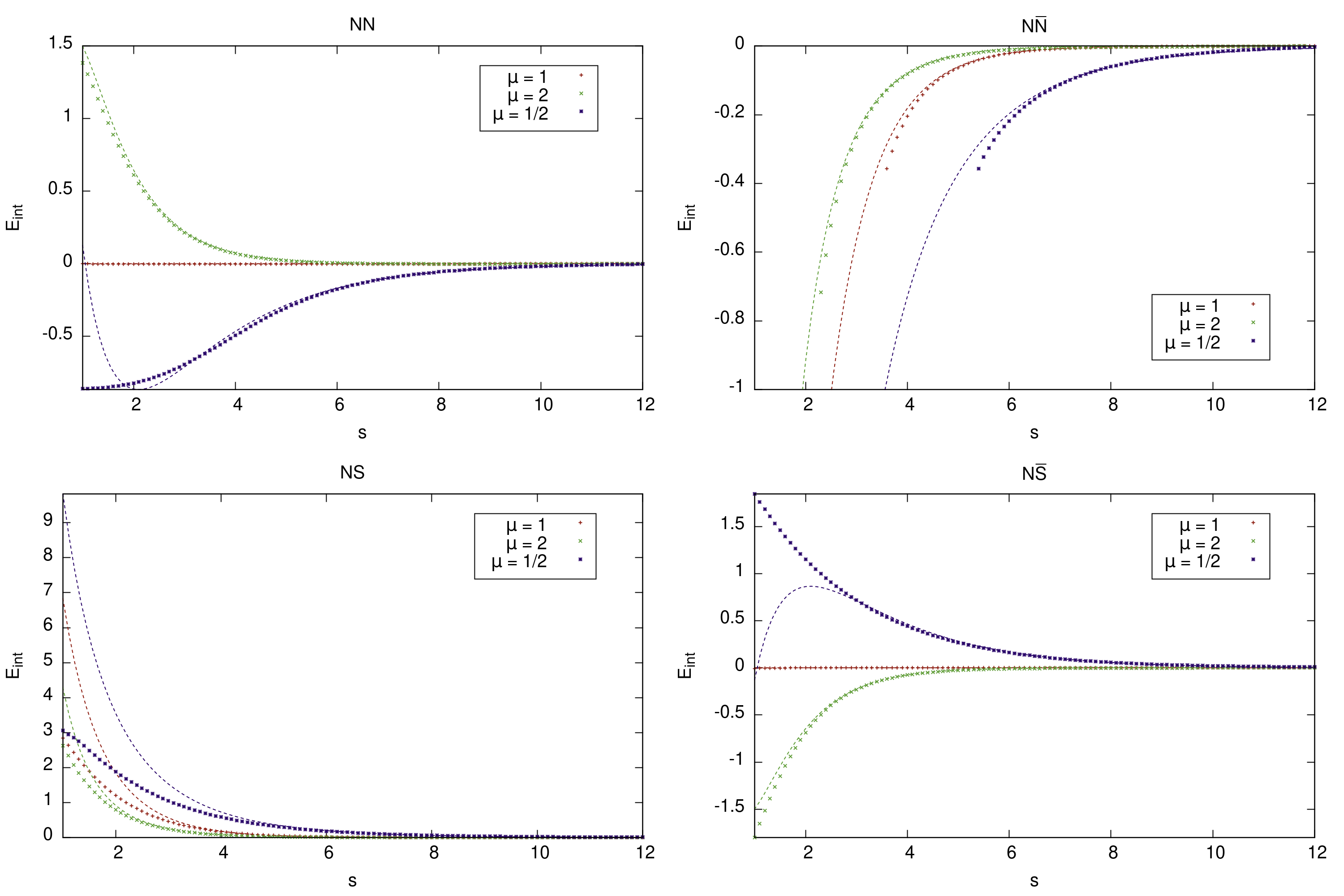}
	\caption{Plot of the interaction energies for different vortex pairs and separations $E_{int} = E - 2 E^1$. The dashed lines are the point vortex approximations given by \eqref{Eint}. Note that the interactions agree with table \ref{table1}. }
	\label{Fig:interactions}
\end{figure}

Interaction energies for any other vortex combination can be computed similarly by modifying the constraint \eqref{constraint} and boundary behaviour of the field configuration appropriately. By symmetry, $NN\equiv\bar{N}\bar{N}\equiv SS\equiv\bar{S}\bar{S}$, $N\bar{N}\equiv S\bar{S}$,
$NS\equiv\bar{N}\bar{S}$ and $N\bar{S}\equiv \bar{N}S$, so only 4 of the 10 distinct (anti)vortex pairs need be considered, and we can, without loss of generality, assume that the left vortex is $N$. The results are depicted in Figure \ref{Fig:interactions}. They match perfectly the predictions of our simple
point vortex model.

\section{Concluding remarks}\news\label{sec:conc}
In this paper we have developed a simple point vortex model of long range interactions between (anti)vortices in the usual Ginzburg-Landau model of competing-order superconductors. The model supports two distinct species of vortex, each with a matching antivortex, and hence there are 10 different (anti)vortex pairs possible. Symmetries reduce this to 4 energetically distinct pairs: $NN$, $N\bar{N}$, $NS$ and $N\bar{S}$. The point vortex model predicts asymptotic formulae for the interaction energy of each of these pairs, as a function of separation, with considerable success. This allows us to make typology like arguments similar to those in the standard single component Ginzburg-Landau model. The qualitative nature of the interactions depends on a single parameter $\mu$, the equivalent of the Ginzburg-Landau parameter in the standard model. If $\mu<1$ (equivalent of type I) the interactions display some counterintuitive features. For example, the interaction between vortices of one species and antivortices of the other is {\em repulsive}.

It would be interesting to study vortex lattices in this model in an applied magnetic field. Although, for $\mu>1$, pure $N$ (or pure $S$) arrays are energetically favoured over $NS$ mixtures, if the state emerges from disorder, presumably some species mixing is inevitable. Some work on vortex lattices has already been done \cite{sarkargan,karmakar2018vortex}, however there is further understanding to be gleaned here, as even understanding the ``type" of the superconductor is subtle. In addition, for $\mu < 1$, while it may be preferable for superconducting domains to form rather than vortices, as in a single band superconductor, these domains can now be $N$ or $S$ domains, which will repel each other, leading to meta-stable states. 

Another possibility is the studying of vortex/anti-vortex bound states when applying a magnetic field. While there is previous work on vortices in superconductors \cite{karmengan,sarkargan,wachtel2015signatures,karmakar2018vortex}, the importance of anti-vortices has been completely ignored in the literature until now.

It would also be interesting to consider specific materials such as $YBCO$. Note that while it it is challenging to actually determine the parameter $\mu$ of a given material, it has been suggested that $YBCO$ \cite{hayward2014angular} exhibits vortices and is type II\cite{sarkargan}. In this model this likely means that $\mu >> 1$ so we have vortex/vortex repulsion for all species.

Finally it would be particularly interesting to consider in detail the effect of adding a small term linear in $\rho$ to the original Ginzburg-Landau theory, breaking the energy degeneracy of the two CDW ground states, the upshot of which is that (after rescaling) the energy density becomes
\beq
{\cal E}=\frac12 D_i\uvec\cdot D_i\uvec+\frac12 B^2+\frac{\mu^2}2(\tau-\evec\cdot\uvec)^2,
\eeq
where $\tau$ is an extra small parameter. This term breaks the symmetry between $N$ and $S$ vortices: if $\tau>0$ then $S$ vortices are slightly more energetically costly than $N$ vortices (and {\it vice versa} if $\tau<0$). Remarkably, when $\mu=1$, the model still enjoys a self-duality structure, and $N$ vortices exert no net force on $S$ antivortices. The basic point-vortex model of intervortex forces is similar to the one developed here, in that a point vortex still consists of a scalar monopole of some charge $q$ and a magnetic dipole of some moment $m$, but these sources induce fields of mass $\sqrt{1-\tau^2}\mu$ and
$\sqrt{1-\tau^2}$, and there is no symmetry relating $q^N$ with $q^S$ or $m^N$ with $m^S$. Introducing a linear term has the effect of increasing the range of intervortex forces, therefore, as well as breaking the degeneracy of $N$ and $S$ vortices.

\subsection*{Acknowledgements}
The authors would like to thank Egor Babaev for useful conversations. This work was supported by the UK Engineering and Physical Sciences Research Council through grant EP/P024688/1. All simulations were run using the Soliton Solver library developed by TW at the University of Leeds.

\bibliographystyle{h-physrev}
\bibliography{bibliography}

\end{document}